\documentclass[runningheads]{llncs}
\usepackage{graphicx}
\usepackage{booktabs}
\usepackage{amsmath}

\usepackage{pifont}% http://ctan.org/pkg/pifont
\newcommand{\cmark}{\ding{51}}%
\newcommand{\xmark}{\ding{55}}%
\begin{document}
\title{Dialogue-to-Video Retrieval}
\author{Chenyang Lyu\and
Manh-Duy Nguyen\and
Van-Tu Ninh\and$^{\thanks{The first three authors contributed equally.}}$\\
Liting Zhou\and
Cathal Gurrin \and
Jennifer Foster}
\authorrunning{C. Lyu et al.}
\institute{School of Computing, Dublin City University, Dublin, Ireland \\ \email{\{chenyang.lyu2,manh.nguyen5,van.ninh2\}@mail.dcu.ie \\ \{liting.zhou,cathal.gurrin,jennifer.foster\}@dcu.ie}\\}

% Van Tu Ninh <van.ninh2@mail.dcu.ie>
% \author{Anonymous}
% \institute{}
% \authorrunning{Anonymous}
%
\maketitle              % typeset the header of the contribution
\begin{abstract}
Recent years have witnessed an increasing amount of dialogue/conversation on the web especially on social media. That inspires the development of dialogue-based retrieval, in which retrieving videos based on dialogue is of increasing interest for recommendation systems. Different from other video retrieval tasks, dialogue-to-video retrieval uses structured queries in the form of user-generated dialogue as the search descriptor.
%which contains structured information from utterance. 
%Therefore, 
We present a novel dialogue-to-video retrieval system, incorporating structured conversational information. Experiments conducted on the AVSD dataset show that our proposed approach using plain-text queries improves over the previous counterpart model by 15.8\% on R@1. Furthermore, our approach using dialogue as a query, improves retrieval performance by 4.2\%, 6.2\%, 8.6\% on R@1, R@5 and R@10 and outperforms the state-of-the-art model by 0.7\%, 3.6\% and 6.0\% on R@1, R@5 and R@10 respectively.

% Contribution:
% \begin{itemize}
%     \item Incorporate audio feature in video retrieval
    
%     \item Use dialogue as search query to retrieve video
    
%     \item Making use of CLIP~\cite{radford2021learning_clip} for video retrieval
% \end{itemize}

\end{abstract}

\section{Introduction}

The aim of a video retrieval system is to find the best matching videos according to queries provided by the users~\cite{mithun2018learning,miech2018learning,liu2019use,dzabraev2021mdmmt,cheng2021improving}. Video retrieval has significant practical value as the vast volume of videos on the web has triggered the need for efficient and effective video search systems. 

In this paper, we focus on improving the performance of video retrieval systems by combining both textual descriptions of the target video with interactive dialogues between users discussing the content of the target video. 

Previous work on video retrieval applied a CNN-based architecture~\cite{lecun1998gradient-lenet,krizhevsky2017imagenet-alexnet,he2016deep-resnet} combined with an RNN network~\cite{bahdanau2015neural-attn} to handle visual features and their time-series information \cite{venugopalan2015sequence,yang2018text2video,anne2017localizing}. Meanwhile, another RNN model was employed to embed a textual description into the same vector space as the video, so that their similarity could be computed in order 
%they could be measured its similarity score 
to perform retrieval \cite{mithun2018learning,yang2018text2video,anne2017localizing}. Due to the huge impact of the transformer architecture \cite{vaswani2017transformer} in both text and image modalities, this network has also been widely applied in the video retrieval research field, obtaining improvements over previous approaches \cite{gabeur2020multi,bain2021frozen,luo2021clip4clip,le2022avseeker,hezel2022efficient}. 

Current video retrieval research, however, mainly focuses on plain text queries such as video captions or descriptions. The need to search videos using queries with complex structures becomes more important when the initial simple text query is ambiguous or not sufficiently well described to find the correct relevant video. Nevertheless, there are only a few studies that focus on this problem~\cite{maeoki2020interactive,madasu2022learning}. Madusa et al.~\cite{madasu2022learning} 
%had a similar approach that 
used a dialogue, a sequence of questions and answers about a video, as a query to perform the retrieval because this sequential structure contains rich and detailed information. Specifically, starting with a simple initial description, a video retrieval model would return a list of matching videos from which a question and its answer were generated to create an extended dialogue. This iterative process continued until the correct video was found. Unlike the model of Maeoki et al.\cite{maeoki2020interactive} which applied a CNN-based encoder and an LSTM~\cite{hochreiter1997long_lstm} to embed data from each modality and to generate questions and answers, Madusa et al's system, \textsc{ViReD}~\cite{madasu2022learning}, applied Video2Sum~\cite{song2021towards} to convert a video into a textual summary which can be used with the initial query to get the generated dialogue with the help of a BART model~\cite{lewis-etal-2020-bart}.

In this paper, we focus on a less-studied aspect of video retrieval: dialogue-to-video retrieval where the search query is a user-generated dialogue that contains structured information from each turn of the dialogue. The need for dialogue-to-video retrieval derives from the increasing amount of online conversations on social media, which inspires the development of effective dialogue-to-video retrieval systems for many purposes, especially recommendation systems~\cite{alamri2019audiovisual,he2021improving,zheng2022MMChat}. Different from general text-to-video retrieval, dialogue-to-video uses user-generated dialogues as the search query to retrieve videos. The dialogue contains user discussion about a certain video, which provides dramatically different information than a plain-text query. This is because during the interaction between users in the dialogue, a discussion similar to the following could happen ``A: \textit{The main character of that movie was involved in a horrible car accident when he was 13.} B: \textit{No, I think you mean another character.}''. Such discussion contains subtle information about the video of interest and thus cannot be treated as a plain-text query.

Therefore, to incorporate the conversational information from dialogues, we propose a novel dialogue-to-video retrieval approach. In our proposed model, we sequentially encode each turn of the dialogue to obtain a dialogue-aware query representation with the purpose of retaining the dialogue information. 
Then we calculate the similarity between this dialogue-aware query representation and individual frames in the video in order to obtain a weighted video representation. 
Finally, we use the video representation to compute an overall similarity score with the dialogue-aware query. To validate the effectiveness of our approach, we conduct dialogue-to-video experiments on a benchmark dataset AVSD~\cite{alamri2019audiovisual}. Experimental results show that our approach achieves significant improvements over previous state-of-the-art models including \textsc{FiT} and \textsc{ViReD}~\cite{maeoki2020interactive,bain2021frozen,madasu2022learning}.

% Introduce the novel parts of the task

% Link the new task with the old relevant tasks in VBS (Textual Known Item Search)

% There're increasingly amount of dialogs/conversations on social media, which creates needs to recommend/retrieve videos to social media users

% Describe our system and experimental results.
% Contributions:

% \input{1-Related-Work}

\section{Methodology}

\begin{figure}
    \centering
    \includegraphics[width=0.9\linewidth]{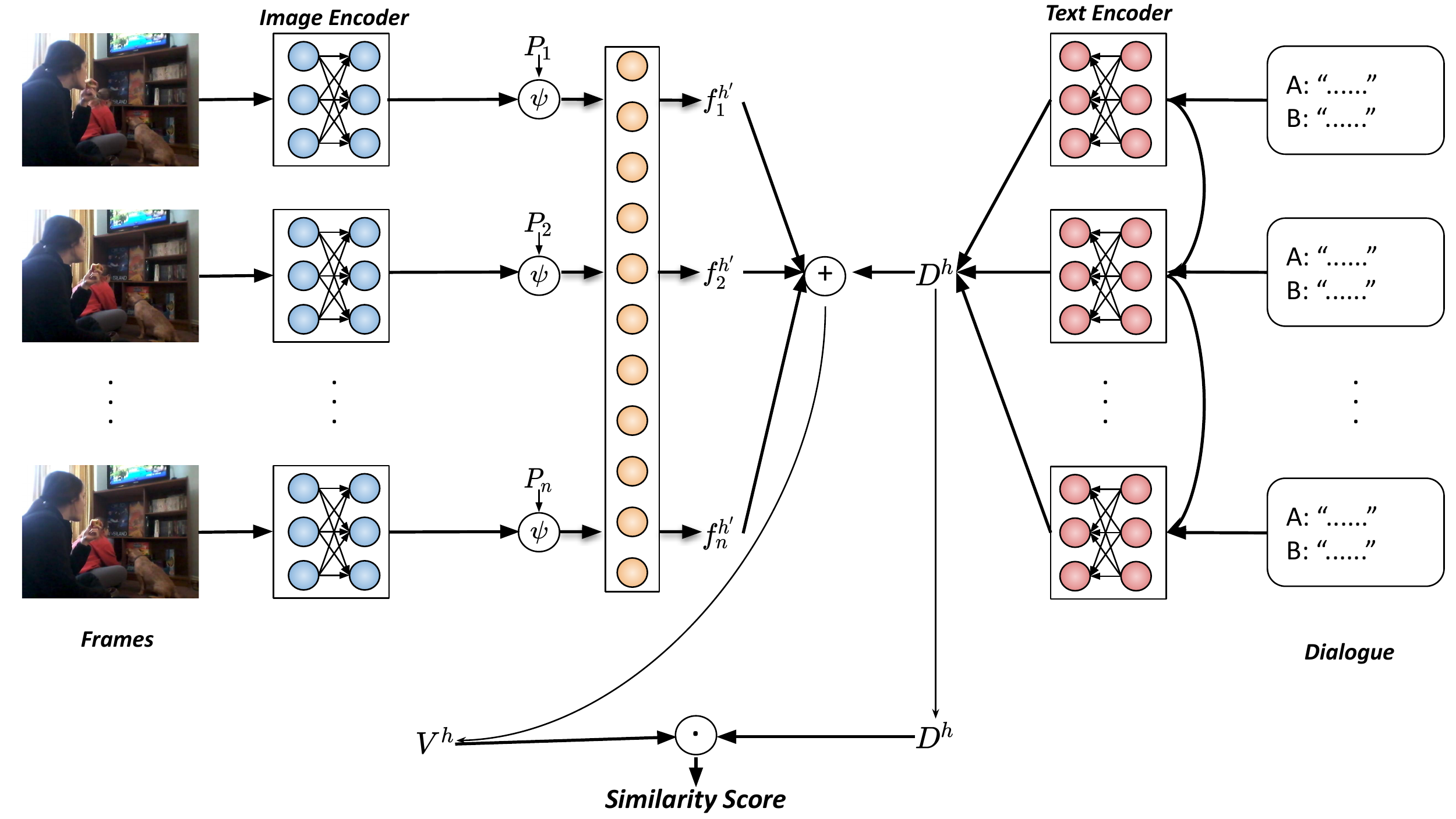}
    \caption{The architecture of our proposed approach.}
    \label{fig:model_architecture}
\end{figure}
In this section, we describe how our dialogue-to-video retrieval system works. Our retrieval system consists of two major components: 1) a \textbf{\textit{temporal-aware video encoder}} responsible for encoding the image frames in video with temporal information. 2) a \textbf{\textit{dialogue-query encoder}} responsible for encoding the dialogue query with conversational information. As shown in Figure~\ref{fig:model_architecture}, our model receives video-query pairs and produces similarity scores. Each video consists of $n$ frames: $V = \{f_{1}, f_{2},......,f_{n}\}$ and each dialogue query is composed of $m$ turns of conversation: $D = \{d_{1},d_{2},......,d_{m}\}$.

In the \textit{video encoder}, we encode each frame $f_{i}$ to its visual representation $f_{i}^{h}$. Then we incorporate temporal information to the corresponding frame representation and feed them into a stacked \textsc{Multi-Head-Attention} module, yielding temporal frame representation $f_{i}^{h^{'}}$. In the \textit{dialogue-query encoder}, we sequentially encode $D$ by letting $d_{i}^{h}=\textsc{Text-Encoder}(d_{i-1}^{h}, d_{i})$ in order to produce a dialogue-history-aware dialogue representation. We then obtain the final dialogue-query representation by fusing all $d_{i}^{h}$: $D^{h} = g(d_{1}^{h},......,d_{m}^{h})$ where $g$ represents our fusion function. After obtaining $D^{h}$, we use it to calculate similarities with each frame $f_{i}^{h^{'}}$, which are then used to obtain a video representation $V^{h}$ based on the weighted summation of all $f_{i}^{h^{'}}$. Finally, we obtain the dialogue-to-video similarity score using the dot-product between $D^{h}$ and $V^{h}$.

\subsection{Temporal-aware Video Encoder}

Our \textit{temporal-aware video encoder}, which is built on Vision Transformer~\cite{dosovitskiy2020image} %based on Transformer~\cite{Vaswani:2017:AYN:3295222.3295349}, 
firstly encodes each frame $f_{i}$ to its visual representation:

\begin{equation}
    f_{i}^{h} = \textsc{Image-Encoder}(f_{i})
\end{equation}

Then we inject the positional information of the corresponding frame in the video to the frame representation and feed it to the \textsc{Multi-Head-Attention} module:

\begin{equation}
    f_{i}^{h'} = \textsc{Multi-Head-Attention}([f_{1}^{p},......,f_{n}^{p}])
\end{equation}
where $f_{i}^{p}$ is the frame representation with positional information $f_{i}^{p}=\psi(f_{i}^{h}, p_{i})$ and $p_{i}$ is the corresponding positional embedding. Practically, we add \textit{absolute} positional embedding vectors to frame representation as in BERT~\cite{bert}: $f_{i}^{p}=f_{i}^{h} + p_{i}$.
Finally, we obtain the temporal-aware video representation $V^{h'} = \{f_{1}^{h'},......,f_{n}^{h'}\}$.

\subsection{Dialogue-query Encoder}

The dialogue-query encoder is responsible for encoding the dialogue-query $D = \{d_{1},d_{2},......,d_{m}\}$:

\begin{equation}
    d_{i}^{h}=\textsc{Text-Encoder}(d_{i-1}^{h}, d_{i})
\end{equation}

where \textsc{Text-Encoder} is a Transformer-based encoder model~\cite{vaswani2017transformer,bert,radford2021learning_clip} in our experiments. Then we fuse all $d_{i}^{h}$ to obtain a dialogue-level representation $D^{h}$ for the dialogue-query:

\begin{equation}
    D^{h} = g(d_{1}^{h},......,d_{m}^{h})
\end{equation}

\subsection{Interaction between Video and Dialogue-query}
To calculate the similarity score between each $V$ and $D$, we firstly compute the similarity scores between dialogue-query $D^{h}$ and each frame $f_{i}^{h'}$. Then we obtain a weighted summation of all frames $f_{i}^{h'}$ as the video representation $V^{h}$:

\begin{equation}
    V^{h} = \sum_{i=1}^{n} c_{i}f_{i}^{h}
\end{equation}

\begin{equation}
    c_{i} = \dfrac{e^{\phi(D^{h}, f_{i}^{h})}}{\sum\limits_{j=1}^{n}e^{\phi(D^{h}, f_{j}^{h})}}
\end{equation}

The final similarity score is obtained by dot-product between $D^{h}$ and $V^{h}$: $s=D^{h}(V^{h})^{T}$
\subsection{Training Objective}

We perform in-batch contrastive learning~\cite{karpukhin2020dense,gao2021simcse}. For a batch of $N$ video-dialogue pairs $\{(V_{1}, D_{1}),......,(V_{N}, D_{N})\}$, the dialogue-to-video and video-to-dialogue match loss are:

\begin{equation}
    L_{d2v} = -\dfrac{1}{N}\sum_{i=1}^{N}\frac{e^{D_{i}^{h}(V_{i}^{h})^{T}}}{\sum\limits_{j=1}^{N}e^{D_{i}^{h}(V_{j}^{h})^{T}}}
\end{equation}

\begin{equation}
    L_{v2d} = -\dfrac{1}{N}\sum_{i=1}^{N}\frac{e^{D_{i}^{h}(V_{i}^{h})^{T}}}{\sum\limits_{j=1}^{N}e^{D_{j}^{h}(V_{i}^{h})^{T}}}
\end{equation}

The overall loss to be minimized during the training process is $L = (L_{d2v}+L_{v2d})/2$.

\section{Experiments}

\subsection{Dataset}
We conduct our experiments on the popular video-dialogue dataset: AVSD~\cite{alamri2019audiovisual}.\footnote{https://video-dialog.com} In AVSD, each video is associated with a 10-round dialogue discussing the content of the corresponding video. We follow the dataset split of AVSD in~\cite{alamri2019audiovisual,maeoki2020interactive}, 7,985 videos for training, 863 videos for validation and 1,000 videos for testing.
\subsection{Training setup}
Our implementation is based on CLIP~\cite{radford2021learning_clip} from Huggingface~\cite{Wolf2019HuggingFacesTS}. CLIP is used to initialize our \textsc{Image-Encoder} and \textsc{Text-Encoder}. For performance and efficiency consideration, we employ ViT-B/16~\cite{radford2021learning_clip} as our image encoder.\footnote{https://openai.com/blog/clip/}. We train our system with a learning rate of $1\times10^{-5}$ for 10 epochs, with a batch size of 16. We use a maximum gradient norm of 1. The optimizer we used is AdamW~\cite{adamw}, for which the $\epsilon$ is set to $1\times10^{-8}$. We perform early stopping when the performance on validation set degrades. We employ R@K, Median Rank and Mean Rank as evaluation metrics ~\cite{alamri2019audiovisual}. Our code is made publicly available.~\footnote{https://github.com/lyuchenyang/Dialogue-to-Video-Retrieval}

\subsection{Results}

\begin{table*}[!htb]
  \centering
  \small
    \caption{Experimental Results on AVSD dataset}
  \begin{tabular}{lcccccccccccc}
    \toprule
    
     && Use Dialogue && R@1 && R@5 && R@10 && MedRank && MeanRank \\
     
    \midrule
    
    LSTM~\cite{maeoki2020interactive} && \cmark && 4.2 && 13.5 && 22.1 && N/A && 119  \\
    \textsc{FiT}~\cite{bain2021frozen} && \xmark && 5.6 && 18.4 && 27.5 && 25 && 95.4  \\
    \textsc{FiT} + Dialogue ~\cite{bain2021frozen} && \cmark && 10.8 && 28.9 && 40 && 18 && 58.7  \\
    % \textsc{ViReD}~\cite{madasu2022learning} && \cmark && 12.0 && 30.5 && 42.1 && 17 && 69.1  \\
    \textsc{ViReD}~\cite{madasu2022learning} && \cmark && 24.9 && 49.0 && 60.8 && 6.0 && 30.3  \\
    \hline
    \textsc{D2V} + Script && \xmark && 21.4 && 45.9 && 57.5 && 9.0 && 39.8  \\
    \textsc{D2V} + Summary && \xmark && 23.4 && 48.5 && 59.1 && 6.0 && 33.5  \\ 
    \textsc{D2V} + Dialogue && \cmark && \textbf{25.6} && \textbf{52.1} && \textbf{65.1} && \textbf{5.0} && \textbf{28.9}  \\

    \bottomrule
    \end{tabular}%
  \label{tbl-0-main-results}
\end{table*}

We present our experimental results on the test set of AVSD~\cite{alamri2019audiovisual} in Table~\ref{tbl-0-main-results}, where we also show the results of recent baseline models including: 1) LSTM~\cite{maeoki2020interactive}, an LSTM-based interactive video retrieval model; 2) \textsc{FiT}~\cite{bain2021frozen}, a Transformer-based text-to-video retrieval model using the video summary as the search query; 3)  \textsc{FiT}~\cite{bain2021frozen} + Dialogue, the \textsc{FiT} model with dialogue in AVSD~\cite{alamri2019audiovisual} as the search query~\footnote{We concatenate all the rounds of dialogue as plain text to serve as the search query.}; 4) \textsc{ViReD}~\cite{madasu2022learning}, a video retrieval system based on \textsc{FiT} and CLIP~\cite{radford2021learning_clip} using the dialogue summary as the initial query and model-generated dialogue as an additional query. 
In Table~\ref{tbl-0-main-results}, our model is named \textsc{D2V}~(\textbf{D}ialogue-\textbf{t}o-\textbf{V}ideo). We also include the results of our system using the the video caption~(script in AVSD dataset) -- \textsc{D2V+Script} -- and the dialogue summary~(summary in AVSD dataset) as the search query -- \textsc{D2V+Summary}. 

The results in Table~\ref{tbl-0-main-results} show that our proposed approach, \textsc{D2V}, achieves superior performance compared to previous models. First, \textsc{D2V+Script} with plain-text video caption input outperforms its counterpart \textsc{FiT} by a large margin~(15.8 R@1 improvement) and even obtains significant improvements (by 10.6 R@1) over \textsc{FiT} using dialogue as input. That shows the effectiveness of our proposed model architecture. Second, \textsc{D2V+Dialogue} significantly outperforms \textsc{D2V+Script} and \textsc{D2V+Summary} by 3.2 R@1 and 2.2 R@1 respectively, which demonstrates the benefit of incorporating dialogue as a search query. The results in Table~\ref{tbl-0-main-results} show that the dialogue does indeed contain important information about the video content and demonstrates the plausibility of using dialogue as a search query.

\begin{figure}[t]
    \centering
    \includegraphics[width=0.9\linewidth]{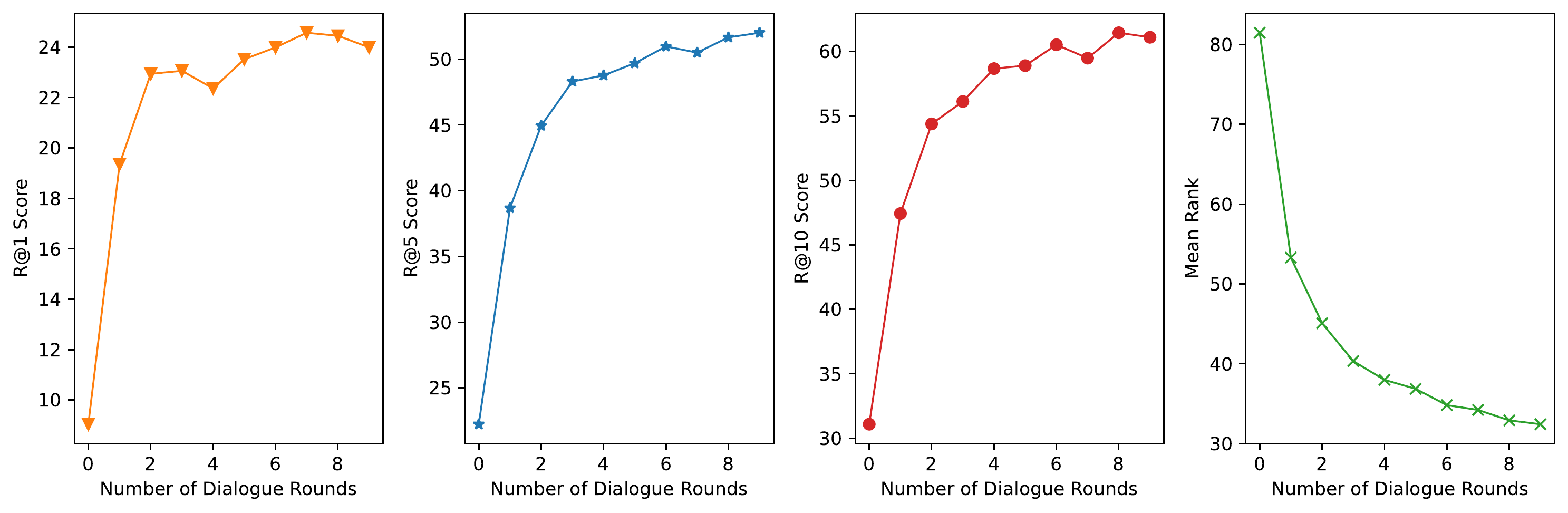}
    \caption{Effect of dialogue rounds}
    \label{fig:effect_of_dialogue_rounds}
\end{figure}

\paragraph{\textbf{Effect of Dialogue Rounds}} 
%Furthermore, 
We investigate the effect of dialogue rounds on the retrieval performance. The results on the validation set of AVSD are shown in Figure~\ref{fig:effect_of_dialogue_rounds}, where we use a varying number of dialogue rounds~(from 1 to 10) when  retrieving videos. 
%Based on the results in Figure~\ref{fig:effect_of_dialogue_rounds},
We observe a consistent improvement with an increasing number of dialogue rounds. The results show that with more rounds of dialogue, we can obtain better retrieval performance. The improvement brought by increasing the dialogue rounds is more significant especially in the early stage~(when using 1 round of dialogue versus 3 rounds).

\section{Conclusion}

In this paper, we proposed a novel dialogue-to-video retrieval model which incorporates conversational information from dialogue-based queries. Experimental results on the AVSD benchmark dataset show that our approach with a plain-text query outperforms previous state-of-the-art models. Moreover, our model using dialogue as a search query yields further improvements in retrieval performance, demonstrating the importance of utilising dialogue information.

\section*{Acknowledgements}
This work was funded by Science Foundation Ireland through the SFI Centre for Research Training in Machine Learning (18/CRT/6183). We thank the reviewers for their helpful comments.

\bibliographystyle{splncs04}
\bibliography{mybibliography}

\begin{thebibliography}{10}
\providecommand{\url}[1]{\texttt{#1}}
\providecommand{\urlprefix}{URL }
\providecommand{\doi}[1]{https://doi.org/#1}

\bibitem{alamri2019audiovisual}
Alamri, H., Cartillier, V., Das, A., Wang, J., Cherian, A., Essa, I., Batra,
  D., Marks, T.K., Hori, C., Anderson, P., Lee, S., Parikh, D.: Audio-visual
  scene-aware dialog. In: Proceedings of the IEEE Conference on Computer Vision
  and Pattern Recognition (2019)

\bibitem{anne2017localizing}
Anne~Hendricks, L., Wang, O., Shechtman, E., Sivic, J., Darrell, T., Russell,
  B.: Localizing moments in video with natural language. In: ICCV (2017)

\bibitem{bahdanau2015neural-attn}
Bahdanau, D., Cho, K.H., Bengio, Y.: Neural machine translation by jointly
  learning to align and translate. In: 3rd International Conference on Learning
  Representations, ICLR 2015 (2015)

\bibitem{bain2021frozen}
Bain, M., Nagrani, A., Varol, G., Zisserman, A.: Frozen in time: A joint video
  and image encoder for end-to-end retrieval. In: IEEE International Conference
  on Computer Vision (2021)

\bibitem{cheng2021improving}
Cheng, X., Lin, H., Wu, X., Yang, F., Shen, D.: Improving video-text retrieval
  by multi-stream corpus alignment and dual softmax loss. arXiv:2109.04290
  (2021)

\bibitem{bert}
Devlin, J., Chang, M.W., Lee, K., Toutanova, K.: {BERT}: Pre-training of deep
  bidirectional transformers for language understanding. In: Proceedings of the
  2019 Conference of the North {A}merican Chapter of the Association for
  Computational Linguistics: Human Language Technologies, Volume 1 (Long and
  Short Papers). pp. 4171--4186. Association for Computational Linguistics,
  Minneapolis, Minnesota (Jun 2019). \doi{10.18653/v1/N19-1423},
  \url{https://www.aclweb.org/anthology/N19-1423}

\bibitem{dosovitskiy2020image}
Dosovitskiy, A., Beyer, L., Kolesnikov, A., Weissenborn, D., Zhai, X.,
  Unterthiner, T., Dehghani, M., Minderer, M., Heigold, G., Gelly, S., et~al.:
  An image is worth 16x16 words: Transformers for image recognition at scale.
  In: International Conference on Learning Representations (2020)

\bibitem{dzabraev2021mdmmt}
Dzabraev, M., Kalashnikov, M., Komkov, S., Petiushko, A.: {MDMMT}: Multidomain
  multimodal transformer for video retrieval. In: CVPR (2021)

\bibitem{gabeur2020multi}
Gabeur, V., Sun, C., Alahari, K., Schmid, C.: Multi-modal transformer for video
  retrieval. In: ECCV (2020)

\bibitem{gao2021simcse}
Gao, T., Yao, X., Chen, D.: {SimCSE}: Simple contrastive learning of sentence
  embeddings. In: Empirical Methods in Natural Language Processing (EMNLP)
  (2021)

\bibitem{he2021improving}
He, F., Wang, Q., Feng, Z., Jiang, W., L{\"u}, Y., Zhu, Y., Tan, X.: Improving
  video retrieval by adaptive margin. In: Proceedings of the 44th International
  ACM SIGIR Conference on Research and Development in Information Retrieval.
  pp. 1359--1368 (2021)

\bibitem{he2016deep-resnet}
He, K., Zhang, X., Ren, S., Sun, J.: Deep residual learning for image
  recognition. In: Proceedings of the IEEE conference on computer vision and
  pattern recognition. pp. 770--778 (2016)

\bibitem{hezel2022efficient}
Hezel, N., Schall, K., Jung, K., Barthel, K.U.: Efficient search and browsing
  of large-scale video collections with vibro. In: International Conference on
  Multimedia Modeling. pp. 487--492. Springer (2022)

\bibitem{hochreiter1997long_lstm}
Hochreiter, S., Schmidhuber, J.: Long short-term memory. Neural computation
  \textbf{9}(8),  1735--1780 (1997)

\bibitem{karpukhin2020dense}
Karpukhin, V., Oğuz, B., Min, S., Lewis, P., Wu, L., Edunov, S., Chen, D.,
  Yih, W.t.: Dense passage retrieval for open-domain question answering. In:
  Empirical Methods in Natural Language Processing (EMNLP) (2020)

\bibitem{krizhevsky2017imagenet-alexnet}
Krizhevsky, A., Sutskever, I., Hinton, G.E.: Imagenet classification with deep
  convolutional neural networks. Communications of the ACM  \textbf{60}(6),
  84--90 (2017)

\bibitem{le2022avseeker}
Le, T.K., Ninh, V.T., Tran, M.K., Healy, G., Gurrin, C., Tran, M.T.: Avseeker:
  an active video retrieval engine at vbs2022. In: International Conference on
  Multimedia Modeling. pp. 537--542. Springer (2022)

\bibitem{lecun1998gradient-lenet}
LeCun, Y., Bottou, L., Bengio, Y., Haffner, P.: Gradient-based learning applied
  to document recognition. Proceedings of the IEEE  \textbf{86}(11),
  2278--2324 (1998)

\bibitem{lewis-etal-2020-bart}
Lewis, M., Liu, Y., Goyal, N., Ghazvininejad, M., Mohamed, A., Levy, O.,
  Stoyanov, V., Zettlemoyer, L.: {BART}: Denoising sequence-to-sequence
  pre-training for natural language generation, translation, and comprehension.
  In: Proceedings of the 58th Annual Meeting of the Association for
  Computational Linguistics. pp. 7871--7880. Association for Computational
  Linguistics, Online (Jul 2020). \doi{10.18653/v1/2020.acl-main.703},
  \url{https://www.aclweb.org/anthology/2020.acl-main.703}

\bibitem{liu2019use}
Liu, Y., Albanie, S., Nagrani, A., Zisserman, A.: Use what you have: Video
  retrieval using representations from collaborative experts. arXiv:1907.13487
  (2019)

\bibitem{adamw}
Loshchilov, I., Hutter, F.: Decoupled weight decay regularization. In:
  International Conference on Learning Representations (2019),
  \url{https://openreview.net/forum?id=Bkg6RiCqY7}

\bibitem{luo2021clip4clip}
Luo, H., Ji, L., Zhong, M., Chen, Y., Lei, W., Duan, N., Li, T.: Clip4clip: An
  empirical study of clip for end to end video clip retrieval and captioning.
  Neurocomputing  (2022)

\bibitem{madasu2022learning}
Madasu, A., Oliva, J., Bertasius, G.: Learning to retrieve videos by asking
  questions. arXiv preprint arXiv:2205.05739  (2022)

\bibitem{maeoki2020interactive}
Maeoki, S., Uehara, K., Harada, T.: Interactive video retrieval with dialog.
  In: Proceedings of the IEEE/CVF Conference on Computer Vision and Pattern
  Recognition Workshops. pp. 952--953 (2020)

\bibitem{miech2018learning}
Miech, A., Laptev, I., Sivic, J.: Learning a text-video embedding from
  incomplete and heterogeneous data. arXiv:1804.02516  (2018)

\bibitem{mithun2018learning}
Mithun, N.C., Li, J., Metze, F., Roy-Chowdhury, A.K.: Learning joint embedding
  with multimodal cues for cross-modal video-text retrieval. In: ICMR (2018)

\bibitem{radford2021learning_clip}
Radford, A., Kim, J.W., Hallacy, C., Ramesh, A., Goh, G., Agarwal, S., Sastry,
  G., Askell, A., Mishkin, P., Clark, J., et~al.: Learning transferable visual
  models from natural language supervision. In: International Conference on
  Machine Learning. pp. 8748--8763. PMLR (2021)

\bibitem{song2021towards}
Song, Y., Chen, S., Jin, Q.: Towards diverse paragraph captioning for untrimmed
  videos. In: Proceedings of the IEEE/CVF Conference on Computer Vision and
  Pattern Recognition. pp. 11245--11254 (2021)

\bibitem{vaswani2017transformer}
Vaswani, A., Shazeer, N., Parmar, N., Uszkoreit, J., Jones, L., Gomez, A.N.,
  Kaiser, L., Polosukhin, I.: Attention is all you need. In: Proceedings of the
  31st International Conference on Neural Information Processing Systems. pp.
  6000--6010. NIPS'17, Curran Associates Inc., USA (2017),
  \url{http://dl.acm.org/citation.cfm?id=3295222.3295349}

\bibitem{venugopalan2015sequence}
Venugopalan, S., Rohrbach, M., Donahue, J., Mooney, R., Darrell, T., Saenko,
  K.: Sequence to sequence-video to text. In: Proceedings of the IEEE
  international conference on computer vision. pp. 4534--4542 (2015)

\bibitem{Wolf2019HuggingFacesTS}
Wolf, T., Debut, L., Sanh, V., Chaumond, J., Delangue, C., Moi, A., Cistac, P.,
  Rault, T., Louf, R., Funtowicz, M., Davison, J., Shleifer, S., von Platen,
  P., Ma, C., Jernite, Y., Plu, J., Xu, C., Le~Scao, T., Gugger, S., Drame, M.,
  Lhoest, Q., Rush, A.: Transformers: State-of-the-art natural language
  processing. In: Proceedings of the 2020 Conference on Empirical Methods in
  Natural Language Processing: System Demonstrations. pp. 38--45. Association
  for Computational Linguistics, Online (Oct 2020).
  \doi{10.18653/v1/2020.emnlp-demos.6},
  \url{https://aclanthology.org/2020.emnlp-demos.6}

\bibitem{yang2018text2video}
Yang, X., Zhang, T., Xu, C.: Text2video: An end-to-end learning framework for
  expressing text with videos. IEEE Transactions on Multimedia  \textbf{20}(9),
   2360--2370 (2018)

\bibitem{zheng2022MMChat}
Zheng, Y., Chen, G., Liu, X., Sun, J.: Mmchat: Multi-modal chat dataset on
  social media. In: Proceedings of The 13th Language Resources and Evaluation
  Conference. European Language Resources Association (2022)

\end{thebibliography}

\end{document}